\documentclass[aps,prb,superscriptaddress,twocolumn,floatfix,showpacs,citeautoscript]{revtex4-2}

\usepackage{float}
\usepackage{amsmath}
\usepackage{amssymb}
\usepackage{amsfonts}
\usepackage{dsfont}
\usepackage{graphicx,graphics}
\usepackage{color}
\usepackage{braket}
\usepackage[dvipsnames]{xcolor}
\usepackage[capitalise,nameinlink]{cleveref}
\crefname{section}{App.}{Apps.}
\usepackage[final]{changes}

\begin{document}
%\title{Optomechanics with quantum microwaves from inelastic Cooper pair tunneling}
\title{Dressed-State Optomechanics in the Few-Photon Regime}

\author{Surangana Sengupta}
\affiliation{Institute for Complex Quantum Systems and IQST, Ulm University, Albert-Einstein-Allee 11, D-89069 Ulm, Germany}
\author{Bj\"orn Kubala}
\affiliation{Institute for Complex Quantum Systems and IQST, Ulm University, Albert-Einstein-Allee 11, D-89069 Ulm, Germany}
\affiliation{German Aerospace Center (DLR), Institute of Quantum Technologies,  Wilhelm-Runge Straße 10, 89081, Ulm Germany}
\author{Joachim Ankerhold}
\affiliation{Institute for Complex Quantum Systems and IQST, Ulm University, Albert-Einstein-Allee 11, D-89069 Ulm, Germany}
\author{Ciprian Padurariu}
\affiliation{Institute for Complex Quantum Systems and IQST, Ulm University, Albert-Einstein-Allee 11, D-89069 Ulm, Germany}
\email{surangana.sen-gupta@uni-ulm.de}
\date{\today}

\begin{abstract}
\noindent Efficient optomechanical cooling typically requires high photon occupancy to maximize cooling power, a constraint that generally limits the degree of coherent quantum control available in the few-photon regime. Here, we investigate this trade-off by considering a strongly nonlinear cavity operated as a discrete quantum system. In the weak-coupling limit, we derive a general connection between the optomechanical damping rate and the cavity's dressed-state manifold. This framework reveals that the damping rate (determined by the population imbalance across dressed states) is directly tunable via the coherent manipulation tools which are standard in circuit quantum electrodynamics. We illustrate this framework using a Josephson photonics architecture, where a dc-biased junction induces a photon blockade that truncates the cavity to an $N$-level system. By sacrificing raw cooling (or heating) power, this platform enables full quantum mechanical control over optomechanical properties, offering a versatile avenue for the quantum manipulation of mechanical modes.
\end{abstract}

\maketitle

\section{Introduction}\label{sec:intro}

Optomechanics provides an important platform for coupling mechanical motion to electromagnetic fields, enabling the cooling and quantum control of massive oscillators for applications in precision sensing \cite{Subpgmasssensing2013,YHeEnhancemassSensing2015,OptmechmassSpec2020,PDjorweEPMassSensing2019,LiUltraHighMassSensing2022} and tests of fundamental quantum mechanics \cite{ChuQAcoustic2017,ArndtMoleculeSupepositon2019,ChuCatstate2023}. The standard approach for cooling mechanical modes to their quantum ground state is resolved-sideband cooling \cite{Kippenbergbasics2007,TeufelGSCOOL2011,PainterGSCOOL2011}. This technique relies on driving an optomechanical cavity red-detuned from resonance to maximize the anti-Stokes process. In this process, photons scattering off the mechanical mode remove phonons from the oscillator, resulting in an effective cooling rate $\Gamma_\text{opt}$ \cite{KippenbergStokesAntiStokes2005}. Conventional cooling requires a high intracavity photon occupation ($n \gg 1$) to achieve a strong optomechanical damping rate. This high-occupation condition allows the system dynamics to be accurately described by a semi-classical approach relating $\Gamma_\text{opt}$ to the cavity's photon number noise spectrum \cite{Clerk_PRL99}.

Experiments have successfully achieved ground-state cooling in various platforms, including optical \cite{PainterGSCOOL2011} and microwave frequency domains \cite{TeufelGSCOOL2011}. However, all standard resolved-sideband cooling protocols face a fundamental trade-off: maximizing the net cooling rate requires a high average photon occupation ($n$), while quantum mechanical control and operation in the truly few-photon regime require $n \sim 1..10$. This conflict restricts coherent, quantum-level manipulation of the cooling process in most current scenarios. Nevertheless, this weak-cooling regime is particularly relevant for platforms where the mechanical mode is replaced by a high-frequency oscillator, such as microwave resonators in radiation-pressure-coupled microwave circuits \cite{Bothner2021a,Bothner2021b}, or bulk acoustic \cite{Renninger2018,Kharel2019, Diamandi2025} and piezoelectric modes \cite{Bochmann2013,Okada2018} in optical systems. Due to their high frequencies, these modes possess inherently low thermal occupations, meaning that even a modest damping rate can significantly influence their final quantum state.

Recent theoretical and experimental efforts, including work on passive Kerr nonlinearities \cite{MetelmanExp2023,MetelmanTheory2025} and our own analysis of active nonlinear driving \cite{sengupta2025}, have primarily focused on operating within the high-$n$ regime (weak nonlinearity) to achieve high damping and suppress backaction heating. Moving beyond this, an important objective is to establish a regime where the cooling dynamics can be accessed and coherently controlled by addressing the cavity as a discrete quantum system.

Here, we develop an intuitive theoretical framework for optomechanical systems using a strongly nonlinear cavity, either intrinsically nonlinear or nonlinearly driven, operated deep in the few-photon quantum regime.

First, we derive a general connection between the optomechanical damping rate and the dressed-state structure of the cavity. In this work, we focus specifically on the weak optomechanical coupling regime. While the few-photon limit often necessitates strong coupling to observe appreciable optomechanical effects, we treat the weak-coupling case as a necessary foundation. This choice allows for a transparent mapping of the optomechanical dynamics onto the dressed-state manifold, providing a rigorous base for future explorations of the strong-coupling regime. By focusing on the dressed-state-resolved regime, where the energy differences between dressed states exceed the cavity broadening, we establish that phonon scattering is primarily driven by transitions between discrete pairs of dressed states. We show that the resulting damping rate is directly proportional to the population imbalance between these states, providing an intuitive link between the cavity’s internal quantum manifold and its optomechanical properties. This dressed-state picture offers a powerful mechanism for quantum coherent control, as the energy differences (which tune the optimal cooling frequency) and the population distribution (which governs the damping rate) can be engineered using the sophisticated control tools typical of modern quantum architectures.

Second, we illustrate this general theory using the nontrivial example of a Josephson-photonics-architecture. Here, a strong nonlinearity is realized via a dc-biased Josephson junction [see Fig.~\ref{fig:circuit}] \cite{Bright_Side,Rolland2019,Menard2022}, which restricts the cavity dynamics to its lowest $N$ levels through a photon blockade. This experimental platform is by now well-established and highly tunable, serving as a promising testbed for coherent control; both the transition frequencies and the steady-state populations are fully tunable using standard circuit quantum electrodynamics techniques.

Our results demonstrate a degree of mechanical control, such as the ability to simultaneously cool and heat different modes or suppress backaction without large detunings, that is fundamentally inaccessible in traditional high-photon-number architectures. Ultimately, this work paves the way for a new generation of hybrid devices where the internal quantum structure of the cavity serves as a versatile tool for mechanical manipulation and, in the future, the coherent control of non-classical mechanical states.

\section{Dressed states optomechanics}
\label{sec:pre}

\subsection{General formulation}
\label{sec:general}

We consider a general optomechanical system described by the Hamiltonian
$\hat{H}= \hat{H}_\text{cav} + \hat{H}_m + \hat{H}_\text{int}$,
where $\hat{H}_\text{cav}$ accounts for the cavity and its drive, potentially including non-linearities of arbitrary complexity. The mechanical mode is governed by $\hat{H}_m = \hbar \omega_m \hat{b}^\dagger \hat{b}$, while the interaction term $\hat{H}_\text{int} = \hbar g_0 (\hat{b}^\dagger + \hat{b})\hat{a}^\dagger \hat{a}$ couples the photon number to the mechanical displacement via the single-photon coupling strength $g_0$.

Dissipative dynamics for the coupled system are described by the Lindblad master equation:
\begin{align} \label{Eqn_Lindblad}
   \frac{d}{dt}\hat{\rho}=-\frac{i}{\hbar}[\hat{H},\hat{\rho}]+\gamma\mathcal{D}[\hat{a}]\hat{\rho} + {}&\gamma_m(\bar{n}_m^\text{th}+1)\mathcal{D}[\hat{b}]\hat{\rho}+\\{}& 
   \gamma_m\bar{n}_m^\text{th}\mathcal{D}[\hat{b}^\dagger]\hat{\rho}\, ,\notag
\end{align}
where $\gamma$ and $\gamma_m$ are the cavity and mechanical decay rates, respectively. We assume the cavity operates in the limit of negligible thermal occupation (e.g., cryogenic temperatures in the case of microwave cavities), while the thermal occupation of the mechanical mode is $\bar{n}_m^\text{th}$. The dissipator is defined as $\mathcal{D}[\hat{L}]\hat{\rho} = \hat{L} \hat{\rho} \hat{L}^\dagger-\frac{1}{2}\{\hat{L}^\dagger\hat{L},\hat{\rho}\}$.

This formulation is applicable to a broad class of systems, including cavities with arbitrary intrinsic nonlinearities or those subject to nonlinear driving, as described by the cavity Hamiltonian $\hat{H}_{\text{cav}}(t)$. We assume a single-mode cavity driven periodically at a fundamental frequency $\omega_d$. Provided the driving amplitude is small relative to $\omega_d$, a rotating-wave approximation yields a time-independent Hamiltonian $\hat{H}_{\text{cav}}$ in the appropriate frame. The eigenstates of this Hamiltonian define the cavity dressed states $\ket{\alpha}$, which satisfy $\hat{H}_{\text{cav}}\ket{\alpha} = \varepsilon_\alpha \ket{\alpha}$. These states constitute the discrete quantum manifold that plays a central role in our theoretical description.

\subsection{Optomechanical damping rate}
\label{sec:opt}

In the few-photon regime, the interaction between mechanical motion and the cavity field is fundamentally shaped by the discrete energy structure of the cavity. For the driven cavity, this structure is defined by the dressed-state energies. While the role of dressed states in optomechanical systems has been characterized in specific hybrid systems featuring embedded two-level emitters or superconducting qubits \cite{Pirkkalainen2015,Viennot2018,Kounalakis2020,Nongthombam2021}, our objective here is to extend this mapping to arbitrary cavity nonlinearities. By doing so, we establish a rigorous theoretical framework for the dressed-state-resolved regime.

In the limit of weak optomechanical coupling $g_0\ll\gamma_m \ll \omega_m$, the influence of the cavity on the mechanical motion is characterized by the optomechanical damping rate $\Gamma_{\text{opt}}$. This rate is obtained using Fermi's Golden Rule and arises from the imbalance in the cavity's photon number fluctuations at the mechanical sidebands \cite{Clerk_PRL99,Clerk_quantum_noise}
\begin{align} \label{Eqn_gamma_opt}
\Gamma_\text{opt} = g_0^2 [ S_{\text{nn}}(\omega_m)-S_{\text{nn}}(-\omega_m)]\,,
\end{align}
where $S_{\text{nn}}(\omega) = \int^\infty_{-\infty} dt e^{i\omega t} \tilde{S}_{nn}(t)$ represents the power spectral density of the photon number fluctuations. The corresponding autocorrelation function, defined as $\tilde{S}_\text{nn}(t) = \langle \hat{a}^\dagger(t)\hat{a}(t)\hat{a}^\dagger(0)\hat{a}(0)\rangle - \langle \hat{a}^\dagger \hat{a} \rangle^2$, is evaluated in the uncoupled limit ($g_0 \to 0$) and captures the intrinsic noise properties of the nonlinearly driven cavity.

To elucidate the connection between the Fermi's Golden Rule expression in Eq.~\ref{Eqn_gamma_opt} and the dressed-state framework, we evaluate the correlation function $\tilde{S}_\text{nn}(t)$ utilizing the regression theorem \cite{breuer2007book} and the basis of dressed states. We begin by defining an auxiliary, unnormalized cavity density operator $\hat{\rho}_n(t)$, which evolves from the initial condition $\hat{\rho}_n(0) = \hat{n}\hat{\rho}(0)$. Here, $\hat{n} = \hat{a}^\dagger \hat{a}$ is the photon number operator and $\hat{\rho}(0)$ denotes the steady-state density matrix of the cavity. The correlation function is then directly related to the dynamics of this auxiliary state via $\tilde{S}_\text{nn}(t) = \text{Tr}\{\hat{n}\hat{\rho}_n(t)\} - \langle \hat{n} \rangle^2$. In this way, the evolution of $\hat{\rho}_n(t)$ provides a transparent path to determining the optomechanical damping rate.

The dynamics of $\hat{\rho}_n(t)$ is governed by the uncoupled master equation ($g_0 \to 0$). In the dressed state basis $\{\ket{\alpha}\}$, where the cavity Hamiltonian is diagonal $({H}_\text{cav})_{\alpha\beta}=\varepsilon_\alpha\delta_{\alpha,\beta}$, the evolution of the matrix elements is
\begin{align}
    (\dot{\rho}_n)_{\alpha\beta}=-i \omega_{\alpha\beta} (\rho_n)_{\alpha\beta}+\gamma \left(\Lambda[\hat{\rho}_n]\right)_{\alpha\beta}\,,
\end{align}
where $\omega_{\alpha\beta} = (\varepsilon_\alpha - \varepsilon_\beta)/\hbar$ are the transition frequencies and $\Lambda[\hat{\rho}_n]$ is the standard Lindblad superoperator applied to $\hat{\rho}_n$, given in the dressed-state basis by 
\begin{align}
      \left(\Lambda[\hat{\rho}_n]\right)_{\alpha\beta}={}&\sum_{\mu,\nu} a_{\alpha \mu} (\rho_n)_{\mu \nu} a^*_{\nu\beta}\\
      {}&-\frac{1}{2} \sum_\zeta \big[n_{\alpha \zeta}(\rho_n)_{\zeta \beta}+(\rho_n)_{\alpha \zeta }n_{\zeta \beta}\big].\notag
\end{align}
Throughout, we use the notation $X_{\alpha\beta}\equiv\braket{\alpha|\hat{X}|\beta}$.

While ${\Lambda}$ generally couples all matrix elements of $\hat{\rho}_n$, the dynamics simplify significantly in the \textit{dressed-state-resolved regime}, characterized by distinct transition frequencies $|\omega_{\alpha\beta}-\omega_{\mu\nu}|\gg \gamma$.

In this dressed-state-resolved regime, each off-diagonal element $(\rho_n)_{\alpha\beta}$ evolves predominantly within a narrow frequency window of width $\sim\gamma$ centered at $\omega_{\alpha\beta}$. While the dissipator $\hat{\Lambda}$ introduces couplings to other matrix elements, these contributions oscillate at frequencies far from the resonance $\omega_{\alpha\beta}$. Consequently, these non-secular terms have an effective amplitude scaling as $\gamma / |\omega_{\alpha\beta} - \omega_{\mu\nu}|$, rendering their impact on the dynamics negligible. Under this approximation, the evolution of the off-diagonal entries decouples
\begin{equation} \label{Eqn_rhon_offdiag}
(\dot{\rho}_n)_{\alpha\beta} \approx (-i\omega_{\alpha\beta} - \tilde{\gamma}_{\alpha\beta}) (\rho_n)_{\alpha\beta},
\end{equation}
where the effective decay rate is $\tilde{\gamma}_{\alpha\beta} = \gamma [ \frac{1}{2}(n_{\alpha\alpha} + n_{\beta\beta}) - a_{\alpha\alpha}a_{\beta\beta}^* ]$.

The resulting solution to \cref{Eqn_rhon_offdiag}, $(\rho_n(t))_{\alpha\beta} = e^{-(i\omega_{\alpha\beta} + \tilde{\gamma}_{\alpha\beta})t} (\rho_n(0))_{\alpha\beta}$, describes a damped oscillation that generates sharp Lorentzian peaks in the noise power spectrum $S_\text{nn}(\omega)$ at the frequencies $\omega_{\alpha\beta}$. These peaks correspond to phonon emission or absorption events associated with the transitions $\ket{\alpha} \leftrightarrow \ket{\beta}$. 

The secular approximation relies on distinct transition frequencies. If certain dressed state transitions are quasi-degenerate ($|\omega_{\alpha\beta} - \omega_{\mu\nu}| \lesssim \gamma$), their corresponding elements remain coupled, forming small linear subsystems that can be solved analytically or semi-analytically. 

It should be noted that our approach is not intended for the conventional scenario of a linear optical cavity under linear driving. In that limit, the Hamiltonian $\hat{H}^{(\text{lin})}_{\text{cav}} = -\hbar\Delta\hat{a}^\dagger\hat{a} + (f/2)(\hat{a}^\dagger + \hat{a})$ possesses an equidistant energy spectrum separated by $\hbar\Delta$. Consequently, all transition frequencies between consecutive levels are identical, leading to fully coupled dynamics. For the linear case, $S_\text{nn}(\omega)$ can be calculated directly \cite{Clerk_PRL99}, without employing a secular approximation.

While the secular approximation effectively decouples the off-diagonal elements, the \textit{diagonal entries} of $\hat{\rho}_n$ are inherently non-oscillatory. In the dressed-state-resolved regime, these elements decouple from the rapidly oscillating off-diagonal terms but remain coupled to one another, evolving via a set of rate equations
\begin{align}
    (\dot{\rho}_n)_{\alpha\alpha}=\gamma \left[\sum_\beta \left(|a_{\alpha\beta}|^2-n_{\alpha\alpha}\delta_{\alpha,\beta}\right)({\rho}_n)_{\beta\beta}\right].
\end{align}
This system of equations can be solved collectively by defining a vector $\vec{v}_n(t)$ composed of the diagonal entries $(\rho_n)_{\alpha\alpha}$. The evolution is then given by $\vec{v}_n(t) = \exp(\gamma \mathbf{A} t) \vec{v}_n(0)$, where the matrix $\mathbf{A}$ has elements $A_{\alpha\beta} = |a_{\alpha\beta}|^2 - n_{\alpha\alpha}\delta_{\alpha\beta}$. Physically, this matrix captures the population transfer between dressed states driven by the cavity dissipation, providing the low-frequency contribution to the correlation function $\tilde{S}_\text{nn}(t)$.

Applying the same secular approximation, but now to the density matrix $\hat{\rho}$ (not to $\hat{\rho}_n$), we find that the steady-state coherences $(\rho(0))_{\alpha \neq \beta}$ are suppressed by a factor of $\gamma/\omega_{\alpha\beta}$ relative to the populations $(\rho(0))_{\alpha\alpha}$. Consequently, the initial condition for the auxiliary matrix elements simplifies to $(\rho_n(0))_{\beta\alpha} = n_{\beta\alpha} P_\alpha$, where $P_\alpha = (\rho(0))_{\alpha\alpha}$ is the steady-state occupation probability of the dressed state $\ket{\alpha}$. Combining these results, the full time-dependence of the correlation function is
\begin{align}\label{Eqn_Snn_t_revised}
\tilde{S}_\text{nn}(t) ={}& \sum_\alpha n_{\alpha\alpha}(\rho_n(t))_{\alpha\alpha} - \langle \hat{n} \rangle^2 \nonumber \\
{}& + \sum_{\alpha \neq \beta} |n_{\alpha\beta}|^2 P_\alpha\, e^{-i\omega_{\alpha\beta}t} e^{-\tilde{\gamma}_{\alpha\beta}t}.
\end{align}
The first line of \cref{Eqn_Snn_t_revised} contributes to low-frequency features in $S_\text{nn}(\omega)$ [cf.~Fig.~\ref{fig:2_columns}(c), shaded region] corresponding to the unresolved regime of optomechanics where $S_\text{nn}$ is probed at $\omega_m\lesssim\gamma$. Here, the detailed dressed state structure is obscured and plays only an indirect role. As illustrated in Fig.~\ref{fig:2_columns}(c) for the Josephson optomechanical device discussed in Sec.~\ref{sec:Joseph_opt}, the agreement between the full quantum calculation and our secular approximation is excellent for $\omega_{\alpha\beta}/\gamma \simeq 10$, with deviations emerging only as the transition frequencies become comparable to the cavity broadening.

In contrast to the first line, the second line of \cref{Eqn_Snn_t_revised} yields a series of Lorentzian peaks at the transition frequencies $\omega_{\alpha\beta}$ [cf.~Fig.~\ref{fig:2_columns}(c)]. These peaks represent a generalization of the optomechanical susceptibility, now shaped by the internal dressed-state structure. Significantly, the weight of each peak is determined by the \textit{steady-state populations} $P_\alpha$. This population dependence breaks the symmetry between positive and negative frequencies, resulting in a net optomechanical damping rate $\Gamma_{\text{opt}}$ with a rich spectral structure
\begin{align} \label{Eqn_g_opt_dressed_revised}
    \Gamma_{\text{opt}}(\omega_m\gg\gamma)=\displaystyle\sum_{\varepsilon_\alpha<\varepsilon_\beta} \left(P_\alpha - P_\beta\right)\Gamma_{\alpha\beta}(\omega_m),
\end{align}
where the single-transition rates are given by
\begin{align} \label{G_plus_minus}
    \Gamma_{\alpha\beta}(\omega_m\gg\gamma)=g_0^2|n_{\alpha\beta}|^2\frac{2\tilde\gamma_{\alpha\beta}}{(\omega_{\beta\alpha}-\omega_m)^2+(\tilde\gamma_{\alpha\beta})^2}.
\end{align}
Equation \eqref{Eqn_g_opt_dressed_revised} constitutes our central result: it explicitly identifies the net cooling rate as a sum over transitions weighted by the \textit{population imbalance} $(P_\alpha - P_\beta)$ between dressed states. The correspondence between \cref{G_plus_minus} and the standard form of Fermi's Golden Rule is now apparent. The transition matrix element $M_{\alpha\beta}$, representing the optomechanical coupling strength between dressed states $\ket{\alpha}$ and $\ket{\beta}$, enters as $|M_{\alpha\beta}|^2 = g_0^2 |n_{\alpha\beta}|^2$. Furthermore, the energy conservation condition of the Golden Rule is expressed through a Lorentzian of width $\tilde{\gamma}_{\alpha\beta}$, reflecting the finite lifetime of the dressed states.

Notably, in the dressed-state-resolved regime where transition peaks are well-separated, the optomechanical response near a resonance is dominated by a single pair of states. In this limit, the damping rate simplifies to
\begin{align} \label{Eqn_g_opt_alpha_beta_revised}
    \Gamma_{\text{opt}}(\omega_m\simeq\omega_{\beta\alpha})\simeq \left(P_\alpha - P_\beta\right)\Gamma_{\alpha\beta}(\omega_m\simeq\omega_{\beta\alpha}),
\end{align}
under the convention $\varepsilon_\alpha < \varepsilon_\beta$. This expression provides an intuitive picture: net cooling or heating is determined by the competition between upward and downward transitions, with the sign of the optomechanical damping dictated solely by the steady-state population distribution across the dressed-state manifold.

Because the optomechanical coupling is assumed to be weak, its effect on the dynamics of the mechanical element does not require a full quantum master equation treatment; instead, it is sufficiently captured by a rate equation including the optomechanical damping $\Gamma_{\text{opt}}$ and the intrinsic mechanical damping $\gamma_m$. In the cooling regime, where the population of the lower-energy dressed state $\ket{\alpha}$ dominates, the steady-state phonon occupancy $\bar{n}_m$ is reduced from its thermal equilibrium value $\bar{n}_m^{\text{th}}$. The rate equation is analogous to standard linear optomechanics \cite{Clerk_PRL99}, which remains valid in this weak-coupling limit, leading to
\begin{align} \label{Eqn_nm}
    \bar{n}_m = \frac{\Gamma_{\text{opt}}\bar{n}_m^{\text{r}}+\gamma_m \bar{n}_m^{\text{th}}}{\Gamma_{\text{opt}}+ \gamma_m}.  
\end{align}
The term $\bar{n}_m^{\text{r}}$ reflects the residual heating or quantum backaction limit. It represents the minimum occupancy achievable when the mechanical resonator is coupled solely to the cavity ($\gamma_m \to 0$). Near a specific transition, this term takes the remarkably simple form
\begin{align} \label{Eqn_nmr}
    \bar{n}_m^{\text{r}}(\omega_m\simeq\omega_{\beta\alpha})\simeq P_\beta/(P_\alpha-P_\beta),  
\end{align}
where we again assume $\varepsilon_\alpha < \varepsilon_\beta$. We note that the dressed-state-resolved regime allows for cooling toward the quantum ground state, provided the system is prepared such that the higher-energy dressed state has negligible population ($P_\beta\to 0$). However, in practice, cooling to the quantum ground state also requires $\Gamma_{\text{opt}}\gg\gamma_m \bar{n}_m^{\text{th}}$, when the optomechanical damping rate overcomes the influx of thermal excitations. This constraint presents a considerable experimental challenge, which motivates further study of the dressed-state-resolved regime in the limit of strong optomechanical coupling. However, such a regime is outside the scope of the present study and will be addressed elsewhere.

Since the transition frequencies $\omega_{\alpha\beta}$, which dictate the optimal damping conditions, and the populations $P_\alpha$, which determine the damping strength, are both governed by the cavity Hamiltonian, they offer significant degrees of freedom for control. These quantities can be engineered via circuit design and dynamically manipulated using standard tools from circuit quantum electrodynamics. This adaptability allows for the precise tuning of the dressed-state manifold to optimize optomechanical performance at several frequencies simultaneously. This approach offers an alternative to conventional methods to address multiple mechanical modes [cf. Sec. IX in Ref.~\cite{Aspelmeyer2014}], which typically require either a ``bad" cavity, where all modes fall within a broad linewidth, or complex multi-frequency driving schemes. By utilizing a single nonlinear cavity mode under monochromatic driving, the engineered dressed-state manifold offers a more scalable multi-mode platform, suggesting promising directions for mechanical lasing [cf. Sec. VIII in Ref.~\cite{Aspelmeyer2014}] and synchronization \cite{Heinrich2011}. 

\begin{figure}[t] 
\includegraphics[width=0.7
\columnwidth] {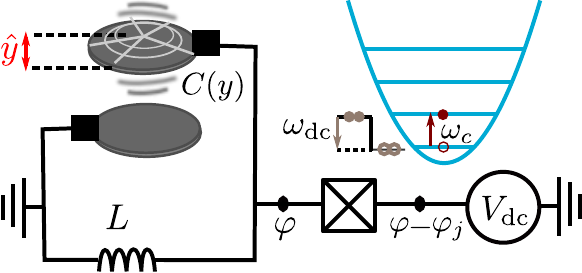}
\caption{{\small(Color online.) Circuit schematic of a dc-biased Josephson junction providing a nonlinear drive to an optomechanical $LC$ resonator integrating a mechanical element as a drumhead capacitor. At the resonance condition $\omega_{\text{dc}} \simeq \omega_c$, energy conservation dictates that each inelastic Cooper pair tunneling event results in the emission of a photon into the cavity mode.
}} 
\label{fig:circuit}
\end{figure}

\section{Josephson optomechanics}
\label{sec:Joseph_opt}

In this section, we illustrate our analysis by considering a superconducting device where dressed states result from an LC cavity driven (nonlinearly) by inelastic Cooper pair tunneling across a voltage-biased Josephson junction, as shown in Fig.~\ref{fig:circuit}. This device concept has been established as a simple, but versatile, source of quantum light, capable of producing single photons \cite{Rolland2019}, entangled pairs \cite{Peugeot2021}, and photon multiplets \cite{Menard2022}. Integrating this device with a mechanical element (e.g. a vibrating capacitive plate) defines a rich optomechanical system. In the regime of weak nonlinearity and high cavity occupation, we have shown that this device facilitates efficient cooling with suppressed residual heating \cite{sengupta2025}. Theoretically, this regime is well-described by a semiclassical approach, in analogy to the theory for standard linear-cavity optomechanics \cite{Clerk_PRL99}. In this work, we focus on the quantum regime where strong driving nonlinearities induce an intrinsic photon blockade, effectively confining the cavity dynamics to a finite manifold of dressed states. We demonstrate that such a device can be feasibly operated in the dressed-state-resolved regime, providing a unique platform to investigate the resulting effects of unusual optomechanical damping.

\subsection{Theoretical model}
\label{sec:Jopt_theo}

\begin{figure*}[t] 
\includegraphics[width=2\columnwidth] {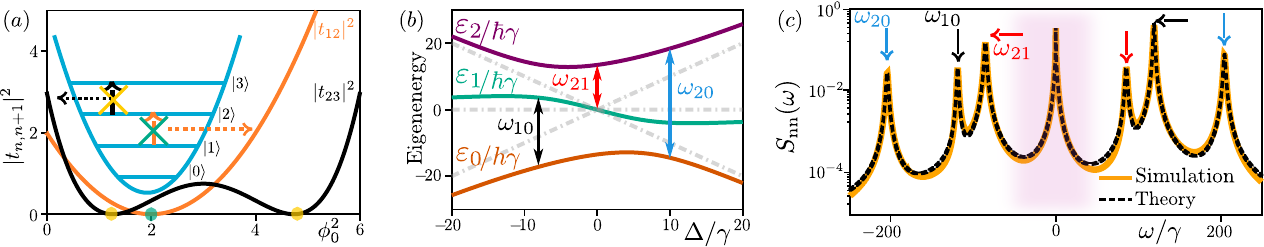}
\caption{{\small(Color online.) Josephson optomechanics in the photon blockade regime. (a) Evidence of photon blockade in the dimensionless transition matrix elements $|t_{n,n+1}|^2 = 4|T_{n,n+1}|^2 / (E_J^*\phi_0)^2$. For the specific values of $\phi_0^2$ marked by green and yellow dots, Fock state transitions $|1\rangle \to |2\rangle$ (orange) and $|2\rangle \to |3\rangle$ (black) are respectively suppressed. (b) Dressed-state eigenenergies for a $3$-level system versus detuning $\Delta$, at $E_J = 300\hbar\gamma$. (c) Photon number spectral density $S_\text{nn}(\omega)$. Arrows identify the transition frequencies between the dressed states of the 3-level manifold ($E_J = 300\hbar\gamma$). The unresolved sideband regime is marked by pink shading.
}} 
\label{fig:2_columns}
\end{figure*} 

%We consider a superconducting circuit, as shown in Fig.~\ref{fig:2_columns}(a), that integrates a mechanical element, exemplified by a vibrating capacitor plate with mechanical degree of freedom $\hat{y}$. 
The Josephson photonics circuit consists of a superconducting microwave cavity (LC resonator) in series with a dc-biased Josephson junction. It integrates a mechanical element, exemplified in Fig.~\ref{fig:circuit} by a vibrating capacitor plate with mechanical degree of freedom $\hat{y}$. The Hamiltonian of the system is given by
%In this manuscript, we present a theoretical description of cavity optomechanics using a superconducting microwave cavity (LC resonator) in series with a dc-biased Josephson junction.  The mechanical degree of freedom $\hat{y}$ can be described by the changing capacitance with respect to the position of the microwave resonator.  (see Fig.\ref{fig:intro}). Given, the LC resonator frequency $\omega_c$,  mechanical oscillator frequency at $y=0$, $\omega_m$ , and the strength of the coupling between the resonator and mechanical oscillator $G$
% \begin{subequations}
% \begin{align} \label{Eqn_Hamil}
%       \ &\hat{H}= \hat{H}_\text{cav} + \hat{H}_m + \hat{H}_\text{int}; \\
%       \ &\hat{H}_\text{cav} = \hbar  \omega_c \hat{a}^\dagger \hat{a}-E_J \cos(\hat{\varphi}_j),\\
%       \ &\hat{H}_m = \hbar \omega_m \hat{b}^\dagger \hat{b}, \\
%       \ &\hat{H}_\text{int} = G \hat{y}\hat{a}^\dagger \hat{a}, 
%       %+f \cos(\Omega t)\hat{y} 
% \end{align}    
% \end{subequations}
\begin{align} \label{Eqn_Hamil}
\hat{H}= \hat{H}_\text{cav} + \hat{H}_m + \hat{H}_\text{int}
\end{align}
consisting of the microwave cavity circuit $\hat{H}_\text{cav} = \hbar  \omega_c \hat{a}^\dagger \hat{a}-E_J \cos(\hat{\varphi}_j)$ that includes the LC-oscillator and the dc-biased Josephson junction, the mechanical mode $\hat{H}_m = \hbar \omega_m \hat{b}^\dagger \hat{b}$, and the optomechanical interaction term $\hat{H}_\text{int} = G \hat{y}\hat{a}^\dagger \hat{a}$.
Here, $E_J$ and $\varphi_j$ are the Josephson energy and phase across the junction, $\omega_c = 1/\sqrt{LC}$ is the frequency of the LC cavity mode, and $\omega_m$ is the frequency of the phonon mode considered. The optomechanical coupling strength $G$ can be related to the change in capacitance (assumed linear) due to the mechanical motion, $G = (dC/dy)(d\omega_c/dC)$. The mechanical displacement is expressed in terms of phonon operators (creation $\hat{b}^\dagger$ and annihilation $\hat{b}$) as $\hat{y}=y_0(\hat{b}^\dagger +\hat{b})$, with $y_0$ the strength of the zero-point fluctuations of the mechanical oscillator. The resulting single-photon coupling strength is given by $\hbar g_0= G y_0$. Similarly, the phase across the LC resonator is expressed in terms of photon operators $\hat{\varphi}=\phi_0 ( \hat{a}^\dagger +\hat{a})$, with zero-point fluctuations $\phi_0$. %determined by the impedance $Z_R$ of the LC circuit at resonance \cite{QED1989}, $\phi_0=\sqrt{4\pi Z_R/R_K$}, where $R_K = h/e^2$ is the von Klitzing constant.

The  voltages are related by Kirchhoff's law, $V_\text{dc} + V_c - V_j = 0$, where the voltage bias is $V_\text{dc}$, the voltage across the LC resonator is $V_c = (\hbar/2e)(d\varphi/dt)$, and the voltage across the Josephson junction is $V_j = (\hbar/2e)(d\varphi_j/dt)$. As a result, the Josephson phase is given by $\hat{\varphi}_j=\hat{\varphi}+\omega_\text{dc}t$, where $\omega_\text{dc} = (2e/\hbar)V_\text{dc}$. The Hamiltonian of the microwave cavity circuit can therefore be conveniently parameterized by the photon operators $\hat{a}^\dagger$ and $\hat{a}$, and in the limit of low bias voltage and low temperature (compared to the superconducting gap) takes the form of a non-linearly driven oscillator,
\begin{align} \label{Eqn_H_cav}
 \hat{H}_\text{cav} = \hbar  \omega_c \hat{a}^\dagger \hat{a}-E_J \cos\left[\phi_0 ( \hat{a}^\dagger +\hat{a})+\omega_\text{dc}t\right].
\end{align}
Energy dissipation in the microwave cavity and in the mechanical mode is described using the Lindblad master equation \cref{Eqn_Lindblad}.

Near the resonance condition $\omega_{\text{dc}} \simeq \omega_c$, the cavity is driven by inelastic Cooper pair tunneling, a process in which each tunneling event converts the Cooper pair energy $2eV_{\text{dc}}$ into an added cavity photon \cite{Bright_Side}. The effect of this nonlinear drive is made apparent after performing a rotating wave approximation, which is well-justified in the standard regime $\omega_{\text{dc}}, \omega_c \gg \gamma \gg \gamma_m$ characteristic of most optomechanical platforms.
We follow the approach in Refs.~\cite{Gramich2013,SemiClassicsAndrew2013} and express the cavity Hamiltonian $\hat{H}_\text{cav}$ in a frame rotating with the frequency corresponding to the voltage bias $\omega_\text{dc}$ using the unitary transformation $\hat{U}=e^{-i\omega_{dc} \hat{a}^\dagger\hat{a} t} $. Neglecting all time-dependent terms in this rotating frame, the nonlinear drive takes the form of an operator whose eigenvalues can be formulated in terms of Laguerre polynomials \cite{Clerk2016,Simon2015} leading to the following effective Hamiltonian that describes the nonlinearly driven cavity, 
\begin{align} \label{Eqn_Hamil_Laguerre}
      \hat{H}_\text{cav}=-\hbar \Delta \hat{a}^\dagger \hat{a} +\frac{E_J^* }{2}\phi_0\left[\vphantom{\frac{E_J}{2}} i\hat{a}^\dagger\hat{B}_1+\text{h.c.}\right]
\end{align}
where the detuning is given by $\Delta=\omega_\text{dc}-\omega_c$ and the effective driving strength is $E_J^*=E_J \exp{(-\phi_0^2/2)}$. The driving strength can be adjusted in such devices \cite{Bright_Side} by employing a flux-tunable Josephson junction in a superconducting quantum interference device geometry \cite{Jaklevic1964}, which allows for the magnetic modulation of $E_J$.
Crucially, $\hat{B}_1$ encodes the nonlinearity as follows,
$$\hat{B}_1=\sum_{n=0}^{\infty} \frac{1}{1+n} L_n^{(1)} (\phi_0^2) \ket{n}\bra{n}.$$
While $\hat{B}_1$ is diagonal in the Fock basis, its dependence on the Fock index $n$ modifies the effective coupling between consecutive Fock states $\ket{n}$ and $\ket{n+1}$. The strength of this nonlinearity is determined by the combined effect of zero-point fluctuations $\phi_0$ of the cavity phase and Fock index $n$, as can be seen from the expansion 
$$\frac{1}{1+n}L_n^{(1)} (\phi_0^2)=1-\frac{n\phi_0^2}{2}+\mathcal{O}(n^2\phi_0^4).$$
For small $\phi_0$, the nonlinear correction $n\phi_0^2$ only becomes significant at high $n$ scaling as $\sim \phi_0^{-2}$. In this limit, the cavity occupation $\langle\hat{n}\rangle$ grows quadratically with drive strength until it reaches a saturation threshold \cite{SemiClassicsAndrew2013}. While this high-photon occupation regime is well-described by semiclassical methods (where $\hat{B}_1$ can be rewritten using a Bessel function) \cite{SemiClassicsAndrew2013, sengupta2025}, here we focus on the opposite limit where $\phi_0$ is large. For large $\phi_0$ of order unity, the saturation in the cavity occupation occurs in the \textit{few-photon regime}, placing the system deep in the quantum regime.

Crucially, in Josephson photonics devices, the zero-point fluctuations are determined by the characteristic impedance $Z_R = \sqrt{L/C}$ of the resonator \cite{QED1989}, such that $\phi_0 = \sqrt{4\pi Z_R/R_K}$, where $R_K = h/e^2$ is the von Klitzing constant. Consequently, $\phi_0$ can be customized during circuit design ($\phi_0\simeq 1$ has been realized in Refs.~\cite{Rolland2019,Menard2022}) or adjusted in situ by engineering a tunable cavity inductance (e.g., using magnetic field-tunable high kinetic inductance materials \cite{Annunziata2010,HoEom2012,Borisov2020} or metamaterials \cite{Kim2019}) or a tunable capacitance (e.g., using the electric field-tunable equilibrium position of a suspended capacitor plate as employed in micro- and nano-electromechanical systems \cite{Bachtold2022}). As we demonstrate below, tuning $\phi_0$ results in cavity photon blockade, where the cavity Hilbert space is effectively restricted to an $N$-level system, providing an ideal platform for exploring optomechanics within the dressed-state-resolved regime.

\subsection{Photon blockade in a nonlinearly-driven cavity}

The effects of the nonlinear drive can be quantified by the transitions it induces between Fock states. Using the cavity Hamiltonian in Eq.~\eqref{Eqn_Hamil_Laguerre}, we calculate the matrix elements $T_{n,n+1} = \bra{n+1}\hat{H}_\text{cav}\ket{n}$ for transitions between consecutive Fock states \cite{Simon2015,Simon2017},
\begin{align}
    T_{n,n+1}=& i\: \frac{E_J^*}{2}\phi_0\:\bra{n+1} \hat{a}^\dagger \hat{B}_1\ket{n}\notag\\
    \ =&i\: \frac{E_J^*}{2}\phi_0\: \frac{1}{\sqrt{n+1}}L_n^1 (\phi_0^2).
    \label{Eqn_transition_matrix_elem}
\end{align} 
For specific values of the zero-point fluctuations $\phi_0$ corresponding to the roots of the Laguerre polynomials, $L_n^1(\phi_0^2) = 0$, the transition element $T_{n,n+1}$ vanishes [cf.~Fig.~\ref{fig:2_columns}(a)]. This suppression can be intuitively understood through the Franck-Condon principle \cite{Franck1926}, since the transition amplitude is proportional to the overlap of displaced Fock states, $T_{n,n+1}\propto\bra{n+1} \hat{D}(i\phi_0)\ket{n}$, due to the cosine form of the Josephson term [cf. \cref{Eqn_H_cav}]. For certain $\phi_0$, the displacement operator $\hat{D}(\alpha)= \exp{(\alpha\hat{a}^\dagger-\alpha^*\hat{a})}$ shifts the wave functions such that their spatial overlap is exactly zero, rendering the transition forbidden. Consequently, higher-energy states become inaccessible, truncating the cavity dynamics to a finite $N$-level system. The specific value of $N$ is thus determined by $\phi_0^2$, as summarized in Table~\ref{tab:n_levels}.

This photon blockade is robust against optomechanical coupling because the interaction depends only on the photon number operator $\hat{n}$, which is diagonal in the Fock basis and thus does not drive transitions across the blockade. Furthermore, this truncation remains valid in the presence of dissipation, provided the cavity environment is sufficiently cold ($k_B T \ll \hbar \omega_c$), a condition typically met in microwave-regime experiments \cite{Bright_Side}.

\begin{figure*}[t]
\includegraphics[width=2\columnwidth]{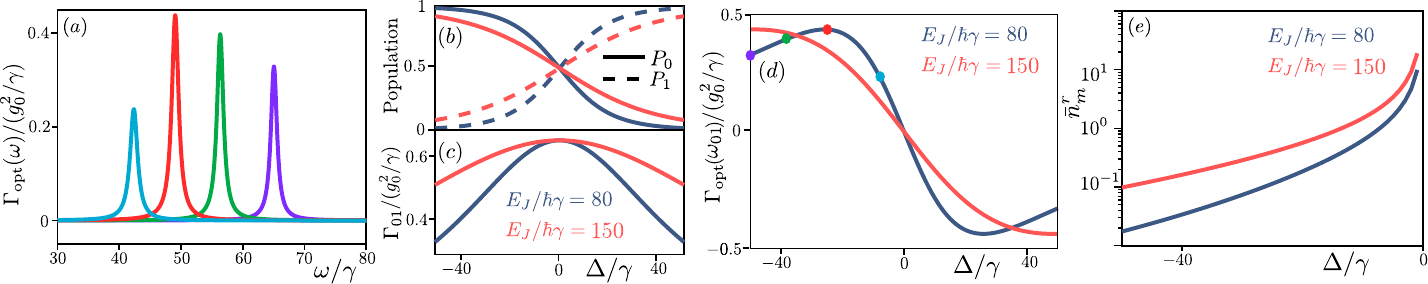}
\caption{\small{(Color online) Optomechanics using an effective $2$-level cavity. (a) Optomechanical damping rate $\Gamma_{\text{opt}}(\omega)$ in the sideband resolved regime $\omega\gg\gamma$. Curves correspond to detunings $\Delta/\gamma = \{-8, -26, -38, -50\}$ (left to right, colored from light blue to purple) at $E_J = 80\hbar\gamma$, with peaks centered at the dressed-state transition $\omega_{10}$. (b) Steady-state populations of the dressed states $\ket{\tilde{0}}$ ($P_0$, dashed) and $\ket{\tilde{1}}$ ($P_1$, solid) versus detuning $\Delta$. Colored curves correspond to different driving amplitudes $E_J/\hbar\gamma=80$ (blue), $150$ (orange). (c) Single-transition rate $\Gamma_{01}(\omega_{10})$ versus detuning $\Delta$. (d) Maximum optomechanical damping rate $\Gamma_{\text{opt}}(\omega_{10})$ plotted versus detuning $\Delta$. Colored round markers correspond to the specific detuning values illustrated in (a). (e) Residual heating phonon number $\bar{n}_m^{\text{r}}$ (logarithmic scale) plotted versus detuning $\Delta$ (linear scale), showing a strong suppression of this unwanted backaction at large detuning $|\Delta|\gg\gamma$.}}
\label{fig:2_LEVEL_g_opt_n_res}
\end{figure*}

%The transition element can be described with an intuitive geometrical picture in terms of the overlap between shifted Fock states of a harmonic oscillator. This becomes obvious when the transition matrix elements are written in terms of the displacement operator $T_{n,n+1}\propto\bra{n+1} D(i\phi_0)\ket{n}$ with $D(\alpha)= \exp{(\alpha\hat{a}^\dagger-\alpha^*a)}$. For certain values of $\phi_0$ this operator shifts the electronic states in a way that the wavefunctions of the two states do no overlap. Therefore owing to the Franck-Condon principle \cite{Franck1926}, fluorescence or absorption is not allowed between such states.

%\begin{figure}[t]
%\begin{center}
%\includegraphics[width=1\columnwidth]{Paper/Figures/N_eJ_TmATRIX.pdf}
%\caption{ \small(Color online.) {(a) Saturation of steady-state photons. Cavity photons in the steady-state plotted as function of $E_J^*$ for $\phi_0=0.5, 0.7, 0.9$ (green, blue and purple lines respectively). The x-axis is in units of $\hbar\gamma=1$. (b) Transition Matrix Elements. The absolute square of the transition matrix elements from $\ket{1} \longrightarrow \ket{2}$ and $\ket{2}\longrightarrow \ket{3}$ is calculated for increasing $\phi_0^2$. The special values of $\phi_0^2$ where the transition is blocked is marked by a blue star at $\phi_0^2=2$ for $N=2$ and by pink stars at $\phi^2=3\pm\sqrt{3}$ for $N=3$. }}
%\label{fig:N_VS_eJ_and_T_n_phi}
%\end{center}
%\end{figure}

\begin{table}[t]
    \centering
     \begin{tabular}{ |c|c|c| } 
\hline
$N$-level & $\phi_0^2$ & Transition blocked \\
\hline
2 & 2 & $\ket{1}\longrightarrow\ket{2}$ \\ 
3 & $3-\sqrt{3}$, $3+\sqrt{3}$ &  $\ket{2}\longrightarrow\ket{3}$\\ 
4 & $0.936$, $3.31$, $7.76$ &  $\ket{3}\longrightarrow\ket{4}$\\ 
5 & $0.743$, $2.57$, $5.73$, $11$ &  $\ket{4}\longrightarrow\ket{5}$\\
6 & $0.617$, $2.11$, $4.61$, $8.40$, $14.3$ &  $\ket{5}\longrightarrow\ket{6}$\\
\hline
\end{tabular}
   \caption{\small{\textbf{Engineered $N$-level cavity.} The different values of $\phi_0^2$ lead to blocking of a certain transitions allowing for the desired $N$-level cavity.}}
    \label{tab:n_levels}
\end{table}

In the following, we focus on these specific cases where the cavity is described by an effective $N$-level Hamiltonian $\hat{H}_{\text{cav},N}$ in the rotating frame 
\begin{align} \label{Eqn_n_n_Hamil}
    \hat{H}_{\text{cav},N}= {}& \sum_{n=0}^{N-1} -\hbar \Delta n\ket{n}\bra{n} +  \\& \notag \sum_{n=0}^{N-2}\left(i\frac{E_J^*}{2}\phi_0 \frac{L_n^{(1)} (\phi_0^2)}{\sqrt{n+1}} \ket{n+1}\bra{n}+\text{h.c.}\right).
\end{align}
The eigenstates of this Hamiltonian constitute the dressed states of the system.

% Transitions between these dressed states can be identified in the emission spectrum of the cavity \cite{QuantumOptics2008},
% \begin{align}\label{Eqn_Spec}
% S(\omega)=\text{Re}\left\{\int dte^{-i\omega t}\braket{\hat{a}^\dagger(t)\hat{a}(0)}_{ss} \right\}    
% \end{align}
% as sideband peaks at frequencies corresponding to all differences between the eigenenergies of the different dressed states.

For an $N$-level cavity, there are $N$ such dressed states with energy eigenvalues $\varepsilon_\alpha$. In general, these states allow for up to $N(N-1)$ distinct transition frequencies $\omega_{\alpha\beta} = (\varepsilon_\alpha - \varepsilon_\beta)/\hbar$, each corresponding to a peak in the noise power spectrum $S_\text{nn}(\omega)$. Due to the symmetry $\omega_{\alpha\beta} = -\omega_{\beta\alpha}$, these peaks appear in pairs symmetric about the origin. Under certain conditions some transition frequencies may coincide [e.g., at zero detuning $\Delta = 0$ for $N=3$, see Fig.~\ref{fig:3_level_panel}(e)]. This leads to overlapping sidebands and the formation of quasi-degenerate subspaces, where the dressed-state-resolved approximation must be handled with the sub-system approach described in Sec.~\ref{sec:opt}.

In the following sections, we validate the theoretical framework developed in Sec.~\ref{sec:pre} by comparing the dressed-state-resolved damping rates calculated using the secular approximation against exact numerical simulations of the full quantum master equation. We specifically examine the cases of a $2$-level and a $3$-level cavity. As we shall demonstrate, these two configurations are qualitatively different: the $2$-level system provides the simplest realization of our framework, while the $3$-level system reveals features generic to cavities with multiple pairs of dressed states.

%\begin{figure}[t]
%\begin{center}
%\includegraphics[width=1\columnwidth]{Paper/Figures/Dressed_state_1.pdf}
%\end{center}
%\caption{ \small{(Color online.) Dressed state analysis for a strongly driven 2-level microwave cavity. (a) The sketch of the energy levels in the Fock basis and the dressed basis. The degenerate states in the Fock basis, split into the dressed states with energy $2\Omega_2$, where $\Omega_2=\sqrt{|T_{01}^2|}$ at $\Delta=0$. (b) The Fourier transformed emission spectrum of the 2-level cavity showing the Mollow triplet at $E_J=30\gamma$ and $\Delta=0$. The position of the Mollow side-bands are at $\pm2\Omega_2$. }}
%        \label{fig:Dressed_state_pic}
%\end{figure}

%\begin{figure}[ht]
%\begin{center}
%\includegraphics[width=1\columnwidth]{Paper/Figures/Spec_and_Snn_2_level_Ej_60_waterfall.pdf}
%\caption{ {\small (Color online.) Emission Spectrum and shot noise spectrum for $N=2$. (a) The spectrum for $N=2$ at finite detuning, $0 \geq \Delta \geq -12$ at $E_J=30 \gamma$ shows a central peak, always at the resonance frequency. The sidebands of the Mollow triplet move outwards as a function of detuning with energy $\Omega_{2d}=\sqrt{\Delta^2+4|T_{01}|^2}$.  The amplitude of all three peaks decrease as we move further away from resonance ($|\Delta|>>0$), as expected. (b) The shot noise spectrum, at $E_J=60 \gamma$ for detunings, $0 \geq \Delta \geq -28$, also shows peaks at resonance frequency and $\pm 2 \Omega_{2d}$.}}
%\label{fig:2_level_Spec_snn_fin_del}
%\end{center}
%\end{figure}

\subsection{Optomechanics using an $N$-level cavity}

We evaluate the validity of our theoretical framework by comparing the analytical results from the dressed-state-resolved regime [Eqs. \eqref{Eqn_Snn_t_revised} and \eqref{Eqn_g_opt_dressed_revised}] against exact numerical solutions of the full quantum master equation \eqref{Eqn_Lindblad}. Specifically, we analyze the photon number spectrum $S_\text{nn}(\omega)$ and the resulting optomechanical damping rate $\Gamma_{\text{opt}}$ for the $N=2$ and $N=3$ cases. These analytical predictions are obtained by diagonalizing the cavity Hamiltonian in the photon blockade regime $\hat{H}_{\text{cav},N}$ in Eq.~\eqref{Eqn_H_cav} and provide a non-trivial example of how the dressed-state structure dictates the optomechanical damping rate.

\subsubsection{Results for a 2-level cavity} \label{sec:2_level}

When the cavity impedance at resonance is tuned such that $\phi_0 = \sqrt{2}$, the transition matrix element $T_{12}$ vanishes, effectively restricting the uncoupled cavity dynamics to the Fock states $\{\ket{0}, \ket{1}\}$. The resulting two-level Hamiltonian in the rotating frame is
\[
\hat{H}_{\text{cav},2}=-\hbar \Delta \ket{1}\bra{1}+ T_{01}\ket{0}\bra{1}+ T_{01}^*\ket{1}\bra{0}
\] 
where $T_{01} = iE_J^*/\sqrt{2}$. The eigenstates of this system are the dressed states $\{\ket{\tilde{0}}, \ket{\tilde{1}}\}$, with corresponding energies $\varepsilon_0 = \hbar\left(-\Delta/2-\Omega_2\right)$ and $\varepsilon_1 = \hbar\left(-\Delta/2+\Omega_2\right)$, where $\hbar\Omega_2 = \sqrt{(\hbar\Delta/2)^2 + |T_{01}|^2}$. In this basis, the Hamiltonian is diagonal
\[
\hat{H}_{\text{cav},2}=-\hbar \Omega_2 \ket{\tilde{0}}\bra{\tilde{0}}+ \hbar\Omega_2\ket{\tilde{1}}\bra{\tilde{1}},
\]
where the energy origin has been shifted by $\hbar\Delta/2$ for convenience.

\begin{figure*}[t]
\includegraphics[width=\linewidth]{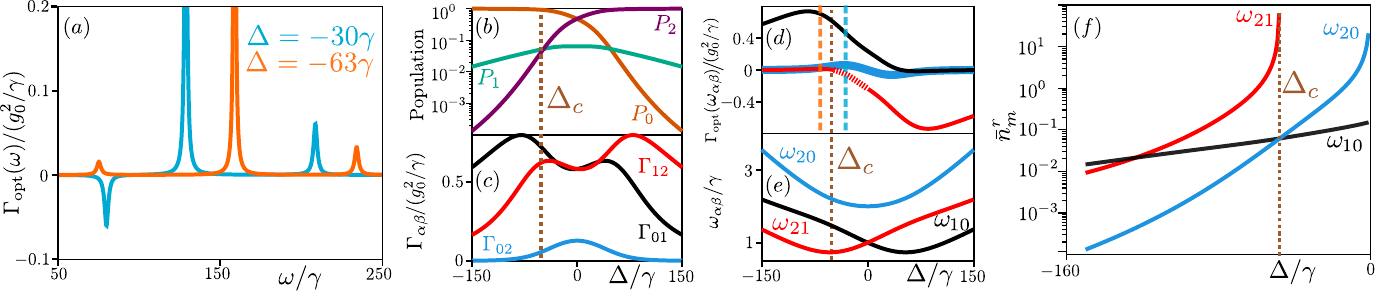}
 \caption{\small{(Color online) Optomechanics using an effective $3$-level cavity. (a) Optomechanical damping rate $\Gamma_{\text{opt}}(\omega)$ for detunings $\Delta/\gamma =-30$ (light blue), and $-63$ (orange). (b) Steady-state populations of dressed states $\ket{\tilde{0}}$ ($P_0$, purple), $\ket{\tilde{1}}$ ($P_1$, light-brown) , and $\ket{\tilde{2}}$ ($P_2$, green)  versus detuning $\Delta$. The vertical dashed line (brown) at $\Delta_c$ marks the population inversion threshold (where $P_2 = P_1$). (c) Single-transition rates $\Gamma_{01}$ (black), $\Gamma_{12}$ (red), and $\Gamma_{02}$ (light blue) vs. detuning. (d) Maximum optomechanical damping rate $\Gamma_{\text{opt}}$ extracted at each resonance: $\omega_{10}$ (black), $\omega_{21}$ (red), and $\omega_{20}$ (light blue). Coloured dashed lines denote the specific detunings illustrated in (a). The dashed region in the curve corresponding to $\omega_{21}$ (red) highlights heating at the red-detuned drive. (e) Transition frequencies $\omega_{\alpha\beta}$ as a function of detuning. (f) Residual heating $\bar{n}_m^{\text{r}}$ extracted for each transition at its respective optimal frequency. All results are calculated for $E_J = 300\hbar\gamma$. }} 
 \label{fig:3_level_panel}
\end{figure*}
The dressed-state-resolved regime is reached when the splitting $\omega_{10} = 2\Omega_2$ significantly exceeds the dissipation rate $\gamma$. This can be achieved even at resonance ($\Delta=0$) by increasing the drive amplitude such that $E_J^* \gg \hbar\gamma$. In this regime, the noise spectrum $S_\text{nn}(\omega)$ exhibits features at frequencies $\pm\omega_{10}$, and the optomechanical damping rate $\Gamma_{\text{opt}}(\omega_m)$ peaks when the mechanical frequency is resonant with the dressed-state transition, $\omega_m = \omega_{10}$ [cf. \cref{fig:2_LEVEL_g_opt_n_res}(a)].

Near this peak, the damping rate is determined by the population imbalance between the dressed states
\begin{align} \label{G_opt_2lvl}
\Gamma_{\text{opt}}(\omega_m\simeq\omega_{10})= (P_0-P_1)\Gamma_{01}(\omega_m\simeq\omega_{10}).
\end{align}
The single-transition rate $\Gamma_{01}$ is defined as
\begin{align} \label{G_2lvl}
    \Gamma_{01}(\omega_m\simeq\omega_{10})=g_0^2|n_{01}|^2\frac{2\tilde\gamma_{01}}{(\omega_{10}-\omega_m)^2+(\tilde\gamma_{01})^2},
\end{align}
with the matrix element $|n_{01}|^2=E^*_J{}^2/(4E^*_J{}^2+2\Delta^2)$ and transition width $\tilde\gamma_{01}= \gamma(3E^*_J{}^2+\Delta^2)/(4E^*_J{}^2+2\Delta^2)$, while the dressed state populations are determined from the steady state of the cavity. 

Physically, Eq.~\eqref{G_opt_2lvl} represents the competition between phonon absorption and emission. When the lower-energy dressed state is more heavily populated ($P_0 > P_1$), absorption dominates, leading to net cooling ($\Gamma_{\text{opt}} > 0$). Similar to conventional linear optomechanics \cite{Clerk_PRL99}, cooling occurs for negative detuning and heating for positive detuning. However, because both the populations and the transition rates depend nonlinearly on the drive $E_J^*$ and detuning $\Delta$ [cf. Fig.~\ref{fig:2_LEVEL_g_opt_n_res}(b) and Fig.~\ref{fig:2_LEVEL_g_opt_n_res}(c)], the resulting damping rate exhibits a complex parameter dependence, as illustrated in Fig.~\ref{fig:2_LEVEL_g_opt_n_res}(d).

Finally, the residual heating $\bar{n}_m^{\text{r}}$ [given by \cref{Eqn_nmr}] quantifies the limits of this cooling process due to unwanted backaction. As shown in Fig.~\ref{fig:2_LEVEL_g_opt_n_res}(e), this unwanted heating $\bar{n}_m^{\text{r}}$ can be suppressed well below unity by increasing the detuning (the same strategy as for linear optomechanics \cite{Clerk_PRL99}), which effectively depopulates the higher-energy dressed state.

In the following, we explore the dressed-state-resolved regime for the slightly more complex case of a 3-level cavity. As we shall see, the addition of a third state introduces multiple transition frequencies and the possibility of population inversion. This leads to a rich landscape where cooling and heating channels coexist at distinct frequencies, offering insight into the multi-level dynamics of the dressed-state manifold.

\subsubsection{Results for a 3-level cavity} \label{sec:3_level}

By tuning the zero-point fluctuations to $\phi_0 = \sqrt{3 - \sqrt{3}}$ (Table \ref{tab:n_levels}), we engineer an effective 3-level cavity restricted to the Fock states $\{\ket{0}, \ket{1}, \ket{2}\}$. The dressed-state energies are found by diagonalizing the Hamiltonian
\begin{align} \label{Eqn_Hamil_3}
    \hat{H}_{\text{cav},3}= {}& -\hbar \Delta \ket{1}\bra{1} -\hbar\; 2\Delta \ket{2}\bra{2} +\\
    {}& \left(T_{01} \ket{1}\bra{0}+T_{12} \ket{2}\bra{1}+\text{h.c.}\right).\notag
\end{align}
where the transition matrix elements are $T_{01} = iE_J^* \phi_0 / 2$ and $T_{12} = iE_J^*[\phi_0 \left(2-\phi_0^2\right)]/{2 \sqrt{2}}$. While the resulting eigenvalues $\varepsilon_\alpha$ have complex analytical forms, they are easily obtained numerically and are plotted in Fig.~\ref{fig:2_columns}(c), where the energy origin has been shifted by $\hbar\Delta$ to ensure that the spectrum is symmetric around the origin in the large-detuning limit [cf. the gray dashed-dotted curves in Fig.~\ref{fig:2_columns}(b), which denote the high-detuning asymptotes of the dressed-state energies].

% The emission spectrum for $N=3$ at resonance $\Delta=0$ consists of the main peak located at $\omega_{dc}$ and two sidebands, as seen in Fig.~\ref{fig:3_level_Spec}(a). The sidebands are positioned at $\omega_{dc} \pm \Omega_3$ and $\omega_{dc} \pm 2\Omega_3$, with $\Omega_3=\sqrt {|T_{01}|^2+|T_{12}|^2}$. At finite detuning $\Delta$, the middle sideband splits into two peaks as shown in Fig.~\ref{fig:3_level_Spec}(b).

% \begin{figure}[t]
% \begin{center}
% \includegraphics[width=1\columnwidth]{Paper/Figures/Spec_Ej_60_3_levels_waterfall_and_det_0.pdf}
% \caption{ {\small (Color online.) Emission Spectrum for $N=3$. (a) The spectrum for $N=3$ at $\Delta=0$, $E_J=60 \gamma$ shows 5 peaks, a central peak, always at the resonance frequency and sidebands at $\pm \Omega_3$ and $\pm 2\Omega_3$, since $\omega_{\psi_\text{gnd}\psi_{e1}}=\omega_{\psi_{e1}\psi_{e2}}=\Omega_3$. (b) The spectrum at finite detuning, $-4 \geq \Delta \geq -16$, splits into more peaks as the degeneracy is lifted, namely $\omega_{\psi_\text{gnd}\psi_{e1}} \neq \omega_{\psi_{e1}\psi_{e2}}$.  }}
% \label{fig:3_level_Spec}
% \end{center}
% \end{figure}

In this $3$-level manifold, the noise spectrum $S_\text{nn}(\omega)$ exhibits three distinct side peaks at the transition frequencies $\omega_{10}$, $\omega_{20}$, and $\omega_{21}$ [cf. Fig.~\ref{fig:2_columns}(c)] and corresponding peaks (cooling) or dips (heating) in the optomechanical damping $\Gamma_\text{opt}$ [cf. Fig.~\ref{fig:3_level_panel}(a)]. A key feature of the $N > 2$ regime is the emergence of a non-trivial population inversion between excited states. As shown in Fig.~\ref{fig:3_level_panel}(b), for moderate negative detunings ($0 > \Delta > -\Delta_c$), the populations $P_1$ and $P_2$ invert such that $P_2 > P_1$, even though the energies maintain the order $\varepsilon_0 < \varepsilon_1 < \varepsilon_2$. This feature is distinct from the population inversion occurring at $\Delta = 0$, which is also present in the $N=2$ case. Non-trivial population inversion arises from the specific driving dynamics within the multi-level manifold, which can favor the steady-state occupation of higher excited states. Consequently, near the transition frequency $\omega_{21}$, the optomechanical damping switches from cooling ($\Delta < -\Delta_c$) to heating ($0>\Delta > -\Delta_c$). Meanwhile, the peaks at $\omega_{10}$ and $\omega_{20}$ continue to provide cooling, as the ground-state population $P_0$ remains dominant. This ability to simultaneously heat a mechanical mode at $\omega_{21}$ while cooling others at $\omega_{10}$ and $\omega_{20}$ using the same cavity mode is a unique signature of the dressed-state-resolved regime.

A similar reversal occurs for positive detuning ($\Delta > 0$). While all three peaks generally lead to heating at large detunings, a population inversion $P_0 > P_1$ at moderate detuning allows for cooling at the frequency $\omega_{10}$, despite the overall heating environment.

The dressed-state populations [Fig.~\ref{fig:3_level_panel}(b)] and transition rates [Fig.~\ref{fig:3_level_panel}(c)] exhibit a non-trivial dependence on detuning and drive amplitude, which is subsequently reflected in the optimized damping rate $\Gamma_{\text{opt}}$ at each transition frequency [Fig.~\ref{fig:3_level_panel}(d)]. Similar to the 2-level system behavior shown in Fig.~\ref{fig:2_LEVEL_g_opt_n_res}(d), these results indicate the existence of an optimal detuning for maximizing the cooling performance of each individual transition. However, it should be noted that the transition frequencies $\omega_{\alpha\beta}$ are themselves functions of the detuning [cf. Fig.~\ref{fig:3_level_panel}(e)]. Consequently, an optimal detuning point implicitly selects a specific transition frequency, which in turn defines the optimal mechanical frequency for cooling in a given experimental implementation.

We note that, as in conventional linear optomechanics, each mechanical frequency can be addressed optimally at either positive or negative detuning by matching the mechanical resonance to a specific transition frequency [cf. Fig.~\ref{fig:3_level_panel}(e)]. The damping rate alternates between heating and cooling upon reversing the sign of the detuning, however, for the population-inverted transition the direction is reversed; here, red detuning leads to heating and blue detuning to cooling.

Furthermore, the 3-level system allows for the suppression of residual heating $\bar{n}_m^{\text{r}}$ without requiring the large detunings typically needed in linear systems. For instance, at moderate negative detuning, the large disparity between $P_1$ and $P_0$ [Fig.~\ref{fig:3_level_panel}(b)] significantly reduces $\bar{n}_m^{\text{r}}$ [Fig.~\ref{fig:3_level_panel}(f)]. In this same regime, the optomechanical damping rate $\Gamma_{\text{opt}}(\omega_{10})$ is also maximized [cf. Fig.~\ref{fig:3_level_panel}(d)], providing optimal conditions for efficient cooling that surpass those found in low-cavity-occupancy linear regimes.

\section{Conclusions} \label{sec:concl}

In this work, we have established a theoretical framework for understanding optomechanical damping and noise in the presence of strong cavity nonlinearities. By shifting the perspective to the dressed-state-resolved regime, we reveal that the cooling dynamics are fundamentally governed by transitions within the cavity's quantum manifold. This approach provides an intuitive link between the optomechanical damping rate and the population imbalance of the dressed states via a result that mirrors Fermi’s Golden Rule while accounting for the finite lifetime of the quantum states.

We illustrated this framework using a Josephson photonics architecture, where the combination of high-impedance circuits and nonlinear driving allows for the precise engineering of a few-level photon manifold. Our results demonstrate that this platform offers unprecedented control: by manipulating the zero-point fluctuations $\phi_0$ and the drive detuning, one can selectively address specific transitions to simultaneously cool or heat different mechanical modes.

While the strong nonlinearity naturally limits the photon occupation, it opens a new door for coherent manipulation of cooling schemes that are inaccessible in linear systems. The ability to suppress residual heating through population engineering, without requiring extreme detunings, highlights the practical advantages of this quantum-resolved approach. 

Furthermore, this cooling scheme offers a direct pathway toward multiplexed optomechanical control. For a target system comprising $M$ mechanical modes, our results suggest a strategy of tuning the nonlinear cavity to a blockade at $N \gtrsim M$ levels, subsequently engineering the drive amplitude and detuning to achieve the desired damping rate for each mode. The utility of this platform may be further extended by incorporating coherent control techniques to manipulate dressed-state populations in real time. Moreover, reaching the single-photon strong-coupling regime would provide additional functionality, enabling even more sophisticated control over the underlying quantum dynamics.

%We have shown how the optomechanical damping can be optimized by reducing the effective fine structure constant of the device $\phi_0^2=4\pi Z_R/R_K$, determined by the impedance of the cavity at resonance. In this semi-classical regime, $\phi_0^2\ll 1$, the residual heating can be reduced only by increasing the detuning and therefore, also reducing the overall damping rate, similarly to conventional optomechanics \cite{Clerk_PRL99}.

% At larger values of the zero-point fluctuations, $\phi_0^2\simeq 1$, we find that the optomechanical damping can be understood as phonon-induced transitions between two discrete dressed states. Each pair of dressed states gives rise to a peak in the frequency dependence of the optomechanical damping rate. The peak weight is proportional to the population difference between the higher and lower energy dressed states. We find that the weights can have different signs, so that both cooling and heating are possible for the same detuning. 

% The residual heating in this dressed states-resolved regime is optimized by reducing the population of the higher energy dressed states. This provides a straight forward method to reduce the residual heating that can also be achieved at small detuning.

% The non-linearly driven microwave cavity presented here is one of the simplest such circuits that can be imagined. As an outlook of this work, we envision designing circuits with multiple nonlinearly driven and possibly intrinsically nonlinear cavity modes. 
Our study of the weak-coupling limit serves as a necessary foundation for this next frontier of strong optomechanical coupling. In that limit, we anticipate that dressed-state transitions will involve the simultaneous exchange of multiple phonons, potentially enabling the simultaneous ground-state cooling of several modes or the generation of multipartite mechanical entanglement. By providing a clear physical picture of the underlying quantum transitions, this work paves the way for a new generation of hybrid devices where the internal structure of the cavity becomes a versatile tool for mechanical control.

\section{Acknowledgements}
We gratefully acknowledge the support of the Deutsche Forschungsgemeinschaft (DFG, German Research Foundation) through
AN336/13-1, AN336/17-1 and AN336/18-1 and the Bundesministerium für
Bildung und Forschung (BMBF) through QSolid.

\bibliography{journalabbreviations,references} % compile four times for best results

\end{document}